\documentclass[
  ,final            
  ]
  {aipproc}

\layoutstyle{8x11double}

\begin{document}

\title[Vlasov simulations of Langmuir Electrostatic Decay]{Vlasov simulations of Langmuir Electrostatic Decay and consequences for Type~III observations.}


\author{P. Henri}{
  address={LESIA, Observatoire de Paris, CNRS, UPMC, Universit\'e Paris Diderot; 5 Place Jules Janssen, 92190 Meudon, France},
  ,altaddress={Dip. Fisica, Universit\`{a} di Pisa, Largo Pontecorvo n.3, 56100 Pisa, Italy},
  email={pierre.henri@obspm.fr}}

\author{F. Califano}{
  address={Dip. Fisica, Universit\`{a} di Pisa, Largo Pontecorvo n.3, 56100 Pisa, Italy},
    ,altaddress={LESIA, Observatoire de Paris, CNRS, UPMC, Universit\'e Paris Diderot; 5 Place Jules Janssen, 92190 Meudon, France}}

\author{C. Briand}{
  address={LESIA, Observatoire de Paris, CNRS, UPMC, Universit\'e Paris Diderot; 5 Place Jules Janssen, 92190 Meudon, France}}

\author{A. Mangeney}{
  address={LESIA, Observatoire de Paris, CNRS, UPMC, Universit\'e Paris Diderot; 5 Place Jules Janssen, 92190 Meudon, France}}

\begin{abstract}
The electrostatic decay enables energy transfer from a finite amplitude Langmuir to a backscattered daughter Langmuir wave and ion acoustic density fluctuations. This mechanism is thought to be a first step for the generation of type III solar radio emissions at twice the plasma frequency.
The electrostatic decay is here investigated through Vlasov-Poisson simulations by considering Langmuir localized wave packets in the case $T_{e} = T_{p}$.
Simulation results are found to be in good agreement with recently reported observations from the STEREO mission of the electrostatic decay of beam-driven Langmuir waves during a type III burst.\end{abstract}

\maketitle


\section{Introduction}

Solar Type~III radio emissions are one of the most prominent features of the meter-decameter ranges of frequency. The emissions show a pronounced drift with time towards lower frequencies. Since the early work of \citet{wild50} and \citet{ginzburg58}, the generally accepted model for such emission is as summarized below. During a flare, high energy electrons (1-100 keV) are expelled from the solar corona and travel along the interplanetary magnetic field lines. The supra-thermal electrons produce a bump on the local electron distribution function generating Langmuir waves via the so-called "bump-on-tail instability". Then, nonlinear wave couplings generate electromagnetic waves at $f_p^-$ (the local electron plasma frequency) or $2f_p^-$. The plasma frequency decreases with the heliocentric distance due to the decrease of the electron density: this is the origin of the time frequency drift characteristic of the Type~III emissions.  \\
Type~III electromagnetic emissions are thought to be produced via two different nonlinear wave-wave couplings.
Through electromagnetic coupling, a mother Langmuir wave $L$ decays into a low frequency $LF$ waves and a transverse electromagnetic wave $T_{f_p^-}$ at the local plasma frequency, observed as Type~III fundamental emission:
\[ L~\rightarrow~T_{f_p^-}+LF  \]
Through electrostatic coupling, known as Langmuir Electrostatic Decay (hereafter LED), the mother Langmuir wave $L$ decays into a low frequency ion acoustic wave $S$ and a daughter Langmuir wave $L'$, which can further non linearly couple with the pump wave to generate a transverse electromagnetic wave $T_{2f_p^-} $ at twice the local plasma frequency, observed as Type~III harmonic emission:
\[
\begin{tabular}{ccc}
	$L~\rightarrow~L'+S$   &  then  & $L'+L~\rightarrow~T_{2f_p^-}$  		
\end{tabular}
\]
We hereafter concentrate on the LED. \\
\citet{Henri&al2009JGR} recently reported direct observations of Langmuir waves decaying into secondary Langmuir waves and acoustic waves during a Type~III solar event, from STEREO/WAVES data. They found that: 
\begin{itemize}
	\item the Doppler-shifted frequencies of the three observed waves satisfy the resonant relations of momentum and energy conservation expected for three-wave coupling 
	\begin{equation} \label{eq:resonantrelations}
		\begin{tabular}{ccc}
			$\omega_L  =  \omega_{L'}+ \omega_S$   &   &
			$\vec{k}_{L}  =  \vec{k}_{L'} + \vec{k}_S$  		
		\end{tabular}
	\end{equation}
	\item a bicoherence analysis confirms the phase coherence of the three waves;
	\item the coupling regions are spatially localized with size of about 2000 Debye lengths.
\end{itemize}
In this former work, the LED threshold and the growth rate of IAW density fluctuations generated by LED were both evaluated from analytical solutions involving a purely monochromatic three-wave coupling \citep{Sagdeev&Galeev1969}. 
However, observations show that: (i) the Langmuir waves are isolated wave packets with a packet width of the order of a few wavelengths; (ii) proton and electron temperatures are known to be close in the solar wind so that IAW associated with the LED should be strongly Landau-damped, this would limit the development of the IAW and thus the LED. 

 The goal of this paper is to study the dynamic of the LED through 1D-1V Vlasov-Poisson simulations
by considering an initial localized Langmuir wave packet, for equal electron and ion temperature. For a better comparison with observed waveform during Type~III bursts, the simulation results are also presented as they would appear if observed by spacecraft instruments.

\section{Description of the model}

LED is here investigated through 1D-1V Vlasov-Poisson simulations in the electrostatic approximation. The description of the code and the numerical scheme can be found in \citet{Mangeney&al2002JCoPh}. 

The Vlasov-Poisson system is solved for the 1D-1V electron and proton distribution function, $f_{e}(x,v,t)$ and $f_{p}(x,u,t)$. The equations are normalized by using the following characteristic electron quantities: the charge $e$, the electron mass $m_{e}$, the electron density $n_{e}$, the plasma (angular) frequency $\omega_{pe} = \sqrt{4 \pi n_{e} e^2 / m_{e}}$, the Debye length $\lambda_{D} = \sqrt{T_{e} / 4 \pi n_{e} e^2}$, the electron thermal velocity $v_{th,e} = \lambda_{D} \omega_{pe} = \sqrt{T_{e} / m_{e}}$ and an electric field $\bar E = m_{e} v_{th,e} \omega_{pe} / e$. Then, the dimensionless equations for each species read:
\begin{equation} \label{eq:vlasovelectrons}
\frac{\partial f_{e}}{\partial t} + v \frac{\partial f_{e}}{\partial x} - (E+E_{ext}) \frac{\partial f_{e}}{\partial v} = 0 
\end{equation}
\begin{equation} \label{eq:vlasovions}
\frac{\partial f_{p}}{\partial t} + u \frac{\partial f_{p}}{\partial x} + \frac{1}{\mu} (E+E_{ext}) \frac{\partial f_{p}}{\partial u} = 0
\end{equation}
\begin{equation} \label{eq:poisson}
\frac{\partial^2 \phi}{\partial x^2} = \int f_{e} dv - \int f_{p} du \\ \mathrm{\ ; \ } \\ E = - \frac{\partial \phi}{\partial x}
\end{equation}
where $v$ (resp. $u$) is the electron (resp. ion) velocity normalized to the electron thermal velocity. Furthermore $\mu=m_{p} / m_{e} =1836$ is the proton-to-electron mass ratio and $\phi$ and $E$ are the self-consistent electric potential and electric field generated by the plasma charge density fluctuations according to Poisson equation (Eq.~\ref{eq:poisson}). Finally, $E_{ext}$ is an "external" driver added to the Vlasov equation that can be switched on or off during the runs. 
The electron (resp. ion) distribution function is discretized in space for $0 \leq x < L_{x}$, with $L_{x} = 10000\ \lambda_{D}$ the total box length, for a resolution of $dx = \lambda_{D}$.
The electron velocity grid ranges over $-5 \leq v/v_{th,e} \leq +5$, with a resolution of $d v = 0.04\ v_{th,e}$. (resp. $-5 \leq u/u_{th,i} \leq +5$, with a resolution of $d u = 0.04\ u_{th,i}$ for the proton velocity grid, where $u_{th,i} = \sqrt{T_{p} / m_{p}}$ is the proton thermal velocity). The temperatures are chosen to be equal $T_{p} = T_{e}$. 

Periodic boundary conditions are used in the spatial direction. The electron and proton distributions functions are initially Maxwellian with respect of velocity, with a random noise in density.

Finally, the initial Langmuir wave packet with the desired wavelength $\lambda_{L}$ and electric field amplitude $E_{L}$ is generated by an "external" electric field $E_{ext}$. The pump $E_{ext}$ oscillates at the expected Langmuir frequency $\omega_{L} = \omega_{pe} + 3/2 (2 \pi \lambda_{D} / \lambda_{L})^2$, with a phase equal to $\omega_{L} t - 2 \pi x / \lambda_{L}$, and is spatially localized with a gaussian-shaped envelop. A Langmuir wave packet propagating only in the forward direction is thus excited. The external electric field $E_{ext}$ is switched off when the generated Langmuir wave reaches the desired amplitude $E_{L}$, typically at $t \sim 300\ \omega_{pe}^{-1}$, much shorter than the LED timescale. The initial Langmuir wave packet then evolves self-consistently. Further details on the forcing can be found in \cite{Henri&alVlasov}.

\section{Langmuir Electrostatic Decay of a localized wavepacket}

In order to compare the simulations results with observations of electric waveforms during Type~III events, Langmuir wavelength and amplitude are set as indicated by solar wind observations.
We choose a Langmuir wave packet with wavelengths centered on $\lambda_{L} = 200\ \lambda_{D}$, a packet width $\Delta = 2000\ \lambda_{D}$, and a maximum initial electric field $E_{L} = 6\ 10^{-2}$. Since the electric field associated with IAW is known to be low, in the following, ion density fluctuations will be used as a tracer for IAW. 

\begin{figure}[h!]
	\begin{tabular}{cc}
	\includegraphics[width=1.05\columnwidth]{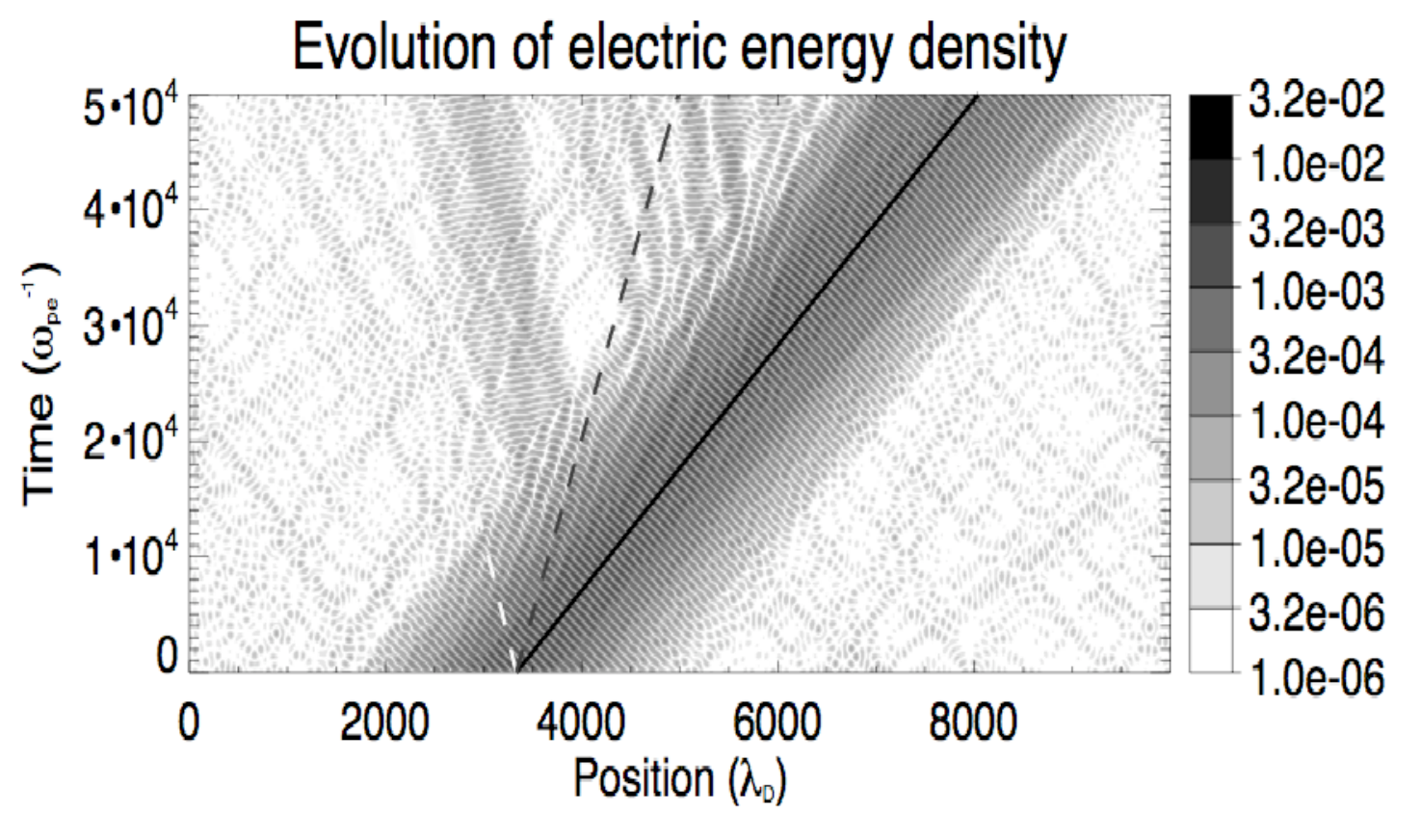} &
	\includegraphics[width=1.05\columnwidth]{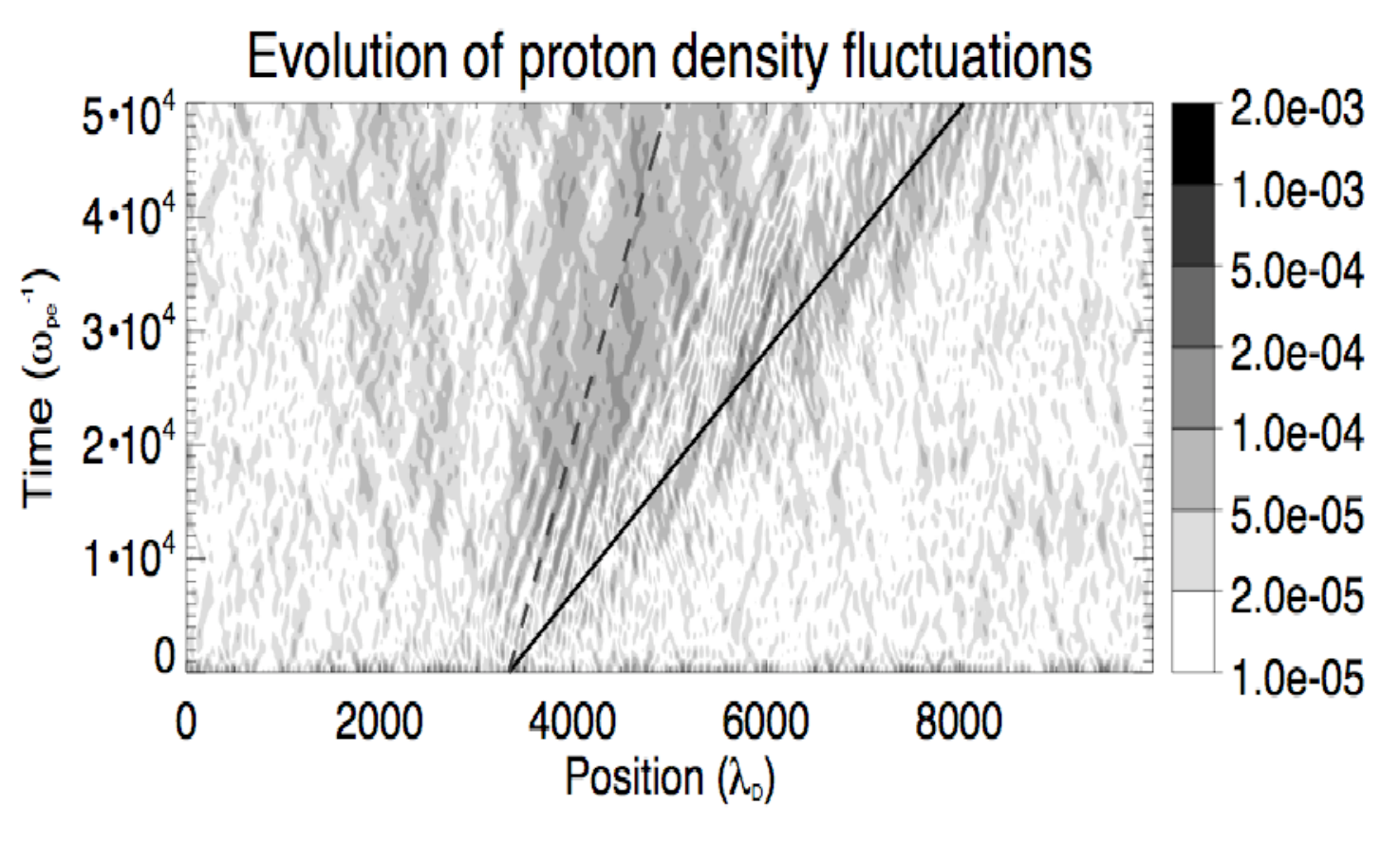} \\
	\end{tabular}
	\caption{Evolution of a localized Langmuir wave packet with initial normalized amplitude $E _{L} = 6\ 10^{-2}$. Top panel: electric density energy $E(x,t)^2/2$. Bottom panel: proton density fluctuations. The expected group velocity of the mother (resp. daughter) Langmuir wave is shown by the black plain line (resp. white dashed line). The ion sound speed is shown by the black dashed line.}
	\label{fig:LEDWavepacketevolution}
\end{figure}

LED is observed in Vlasov simulations for a level of Langmuir electric amplitude typical of those observed in the solar wind during type~III bursts. 
The evolution of the Langmuir wave packet is shown in Fig.~\ref{fig:LEDWavepacketevolution}. 
The electrostatic decay starts at $t \simeq 10^4\ \omega_{pe}^{-1}$.
The mother and daughter Langmuir wave packets can be follow in the top left panel that shows the space-time evolution of the electric density energy $E(x,t)^2/2$. The Langmuir mother wave packet propagates towards the right at its group velocity (black plain line) and emits daughter Langmuir wave packets propagating backward at their own group velocity (white dashed line). 
The bottom left panel shows the temporal evolution of ion density fluctuations during the decay of the Langmuir wave. The fluctuations along the dashed line is a small amplitude artifact of the initial forcing, it does not interact with the electrostatic coupling. 
Instead, ion density fluctuations generated from the Langmuir mother wave packet (along the black solid line) propagate forward at the ion sound speed (slope of the dashed line) with the expected wavelength for the IAW decay product $\lambda_{_{IA}} \simeq 150 \ \lambda_{_{D}}$. They are the decay product.
Note that the IA fluctuations are generated locally where the two Langmuir wave packets beats.  Indeed, IAW density fluctuations are heavily Landau damped, since $T_{e} = T_{p}$, as soon as the waves escape the area where LED occurs. Therefore, the LED-produced IAW is expected to be observed in the solar wind only locally where the mother and daughter Langmuir waves interact. 

In order to compare the simulation results with in-situ waveform observations, we mimic the conditions of observation onboard a spacecraft that would record, on monopole antennas, a decaying Langmuir wave. We thus hereafter introduce in the presentation of simulation results (i) a spacecraft floating potential effect, (ii) a Doppler-shift effect. \\
First, monopole antenna channel are known to be sensitive to both the electric field and the proton density fluctuations, through the fluctuation of the spacecraft potential \citep{Pedersen1995AnGeo, Meyer2007, Kellogg&al2009JGR}. This means that the observed signal is a combination of both electric field and proton density fluctuations. To reproduce this effect in the simulation, we consider an "equivalent signal" $s(t) = E(x,t) + \alpha \ n_{p}(x,t)$, with $\alpha$ the equivalent in the simulation of a calibration parameter that gives in the observations the ratio of the spacecraft floating potential with respect to density fluctuation. This parameter is set to $\alpha = 50$ as an order of magnitude following the prescription of \citet{Kellogg&al2009JGR}. \\
Second, observed waveforms in the spacecraft frame are Doppler-shifted since the plasma is moving at the solar wind speed. A Doppler-shift effect is introduced in the simulation by considering the "equivalent signal" $s(t)$ at position $x(t) = x_{0} + V_{_{S}} t$ where $V_{_{S}}$ is a constant velocity. We choose here $V_{_{S}} = v_{th,e}$ as a first approach. \\
This "equivalent signal" $s(x(t),t)$, obtained from the simulation, represents the signal that would be recorded by a spacecraft crossing such a decaying Langmuir wave packet.

\begin{figure}[h!]
	\begin{tabular}{cc}
	\includegraphics[width=\columnwidth]{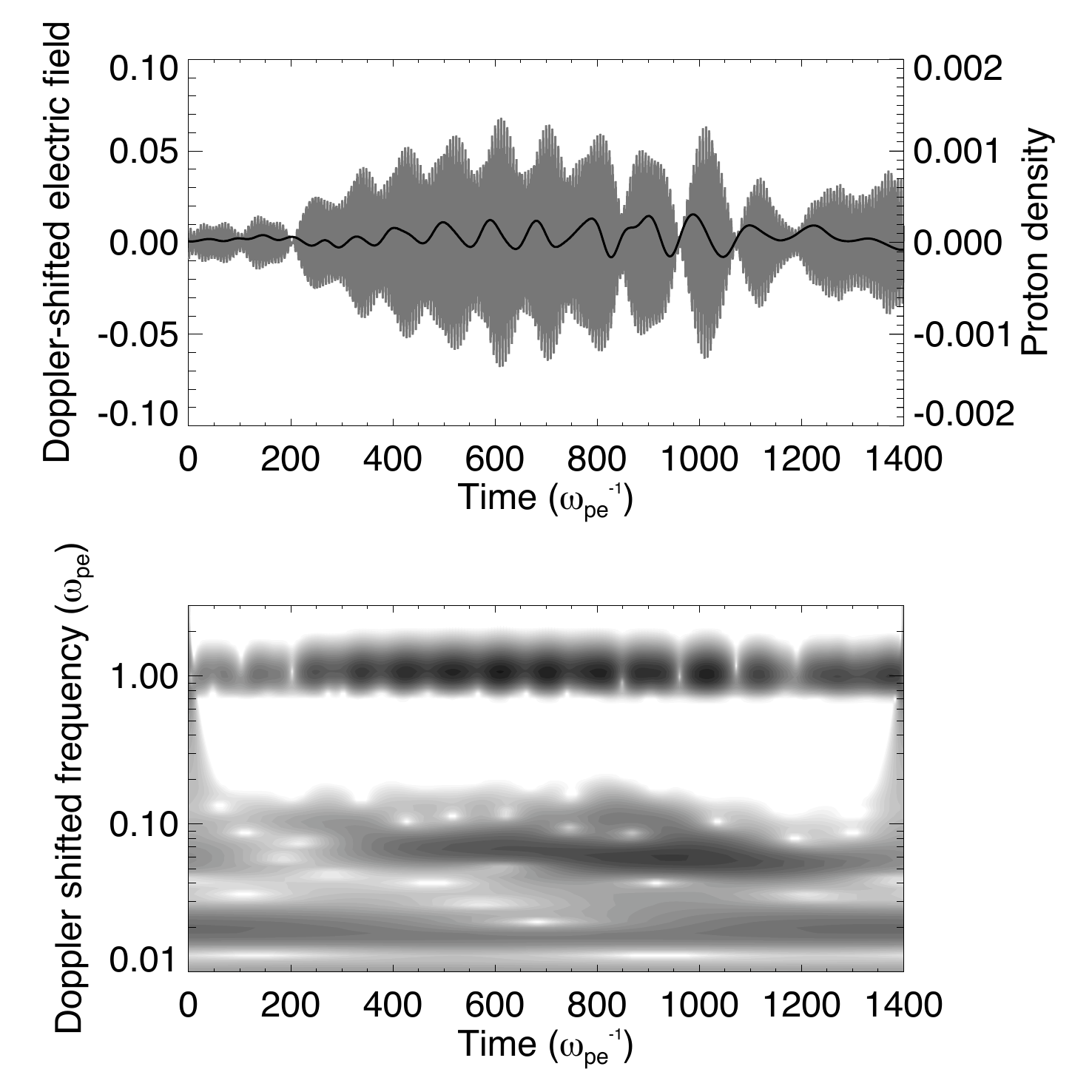} & 
	\includegraphics[width=\columnwidth]{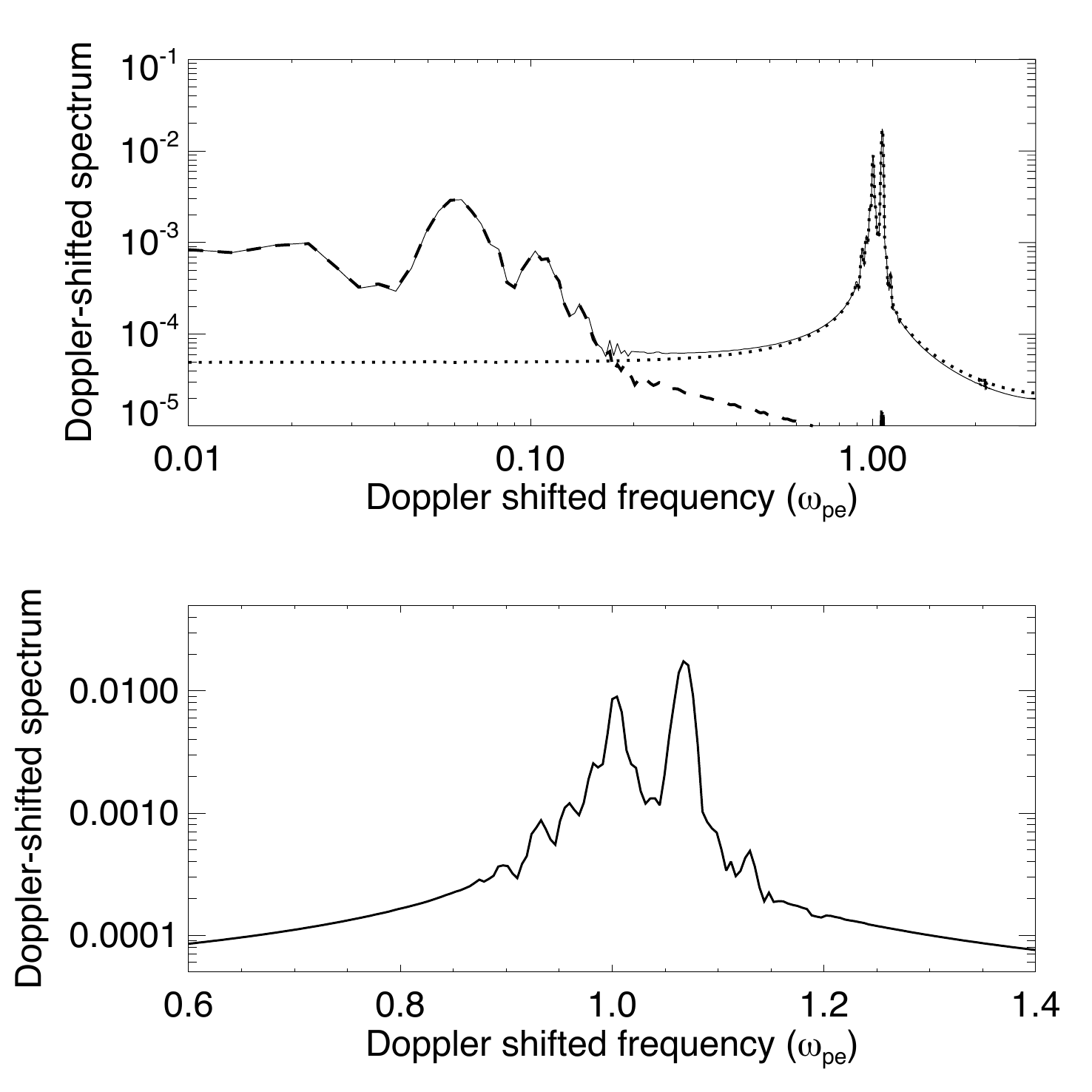} \\
	\end{tabular}
	\caption{Waveform, spectrum and wavelet transform from simulation results, as they would appear when observed by spacecraft instruments when crossing a Langmuir wave under LED. Top left panel: waveform in a moving frame: electric field (grey) and proton density (black). Bottom left panel: Morlet wavelet transform of signal $s(x(t),t)$. Top right panel: corresponding Fourier spectrum (full line) with respective contribution of the density fluctuations (dashed line) and electric field (dotted line). Bottom right panel: zoom on the double peak at the plasma frequency.} \label{fig:LEDWavepacket}
\end{figure}

For an easy confrontation with observed waveforms, simulation results are analyzed and presented in Fig.~\ref{fig:LEDWavepacket} the same way observations have been analyzed and presented by \citet{Henri&al2009JGR}. Both a Doppler-shift effect and the sensitivity to both the electric field and the density fluctuations are taken into account.
The top left panel shows the electric field (grey) and proton density (black line) waveforms from the simulation. Both are plotted in normalized units, as described in the simulation model. 
The left bottom panel shows the wavelet transform of $s(x(t),t)$. 
The simulations show that (i) beat-like modulation at the plasma frequency would be observed, (ii) the IAW signal would be seen at lower frequency centered where the maximum of the beat-like Langmuir signal is observed. This is in full agreement with the observations. The Fourier spectrum of $s(x(t),t)$ is shown in the right top panel, with a zoom at the plasma frequency in the right bottom panel. The frequency of IAW density fluctuations as well as the separation between the frequency peaks of the two Langmuir waves are essentially a Doppler-shift effect. 
The full line represents the total Fourier spectrum. Dashed and dotted lines distinguish the contribution of the ion density fluctuations and electric field respectively. The simulation thus proves that the high frequency part of the observed spectrum is dominated by the response of the antenna to the fluctuations of the electric field, whereas the low frequency part is dominated by the response of the spacecraft potential to density fluctuations.

The waveform and spectrum obtained from Vlasov simulations agrees qualitatively well with observations reported in \citet{Henri&al2009JGR}. However, in order to check whether the level of density fluctuations is quantitatively consistent with previsions from Vlasov simulations, a full calibration of the floating potential of the STEREO spacecraft is needed in order to get a better evaluation of $\alpha$ that fully takes into account the geometry and dimension of the spacecraft. 

\section{Conclusion}

We have reported 1D-1V Vlasov-Poisson simulations of the Langmuir Electrostatic Decay. The simulations have been done in typical solar wind conditions: (i) equal electron and proton temperature, in order to take into account the strong Landau damping of daughter ion acoustic waves, and (ii) localized wave packets, in order to consider the limited interaction time between the mother and daughter waves, each propagating at its own group velocity. \\
Vlasov simulations in typical solar wind conditions (i) shows that the observed level of Langmuir electric field during type~III burst is high enough for Langmuir Electrostatic Decay to start, (ii) reproduce qualitatively the beat-like Langmuir waveforms as well as the spectrum observed during type~III bursts when obtained from monopole antennas. \\
In order to fully confirm the interpretation of the observed Langmuir waveforms in term of the LED of type~III beam-driven Langmuir waves, an effective threshold will be computed from vlasov simulation in typical solar wind conditions \citep{Henri&alVlasov}. \\
We have study here a first step for the generation of electromagnetic
waves like for example Type~III emission. Further Vlasov-Maxwell
simulations must be performed to study the next step of the process and
check the energy transfer from electrostatic to electromagnetic emissions.


\begin{theacknowledgments}
The authors thanks Nicole Meyer-Vernet for useful discussions on spacecraft floating potential. \\
We are grateful to the italian super-computing center CINECA (Bologna) where part of the calculations where performed. We also acknowledge Dr. C. Cavazzoni for discussion on code performance.
\end{theacknowledgments}

\bibliographystyle{aipproc}   


\end{document}